\documentclass[preprint]{aastex}
\newcommand\psr{PSR~B1257+12 }
\newcommand\text[1]{\mbox{#1}}                                               
\newcommand\mref[1]{(\ref{#1})}

\newcommand{\etal}{{\em et al.}\/\/}
\newcommand\pder[2]{\frac{\partial #1}{\partial #2}}

\newcommand\dt{\frac{d \ }{d t}}

\newcommand\br{{\bf r}}
\newcommand\bR{{\bf R}}

\newcommand\bp{{\bf p}}

\newcommand\bP{{\bf P}}

\newcommand\bZ{{\bf Z}}
\newcommand\bzero{{\bf 0}}
\newcommand\cA{{\cal A}}
\newcommand\cB{{\cal B}}
\newcommand\cC{{\cal C}}
\newcommand\cx{{\rm x}}

\begin{document}

\title{Improved timing formula for the PSR B1257+12 planetary system}

\author{Maciej Konacki \altaffilmark{1},
Andrzej J.~Maciejewski  \altaffilmark{2}}
\affil{Toru\'n Centre for Astronomy, 
           Nicolaus Copernicus University,\\
           87-100 Toru\'n, Gagarina 11, Poland}
\author{Alex Wolszczan \altaffilmark{3}}
\affil{Department of Astronomy and Astrophysics, Penn State University \\
University Park, PA 16802, USA \\
Toru\'n Centre for Astronomy, Nicolaus Copernicus University \\
ul. Gagarina 11, 87-100 Toru\' n, Poland}
\altaffiltext{1}{e-mail: kmc@astri.uni.torun.pl}
\altaffiltext{2}{e-mail: maciejka@astri.uni.torun.pl}
\altaffiltext{3}{e-mail: alex@astro.psu.edu}

\begin{abstract}
We present a new analysis of the dynamics of the planetary system around
the pulsar B1257+12. A semi-analytical theory of perturbation between 
terrestrial-mass planets B and C is developed and applied to improve
multi-orbit timing formula for this object. We use numerical simulations
of the pulse arrival times for PSR B1257+12 to demonstrate that our new
timing model can serve as a toll to determine the masses of the two planets.
\end{abstract}

\keywords{celestial mechanics --- planetary systems --- 
          pulsars: individual (PSR B1257+12)} 

\section{Introduction}

The first extra-solar planetary system has been discovered around a 
millisecond pulsar B1257+12 \citep{Wolszczan:92::}. The system consists of 
three planets named ${\cA}, {\cB}$ and ${\cC}$ with planets ${\cB}$ and 
${\cC}$ having the orbits close to a 3:2 commensurability. This circumstance 
allows us to analyze the dynamics of the system beyond the classical Keplerian
approximation. Namely, in such configuration, the gravitational interactions 
of planets ${\cB}$ and ${\cC}$ give rise to observable time variations of  
${\cB}$ and ${\cC}$ orbital elements. It was thoroughly discussed 
by \citet{Rasio:92::} and \citet{Malhotra:93a::} 
\citep[see also][]{Malhotra:92::,Malhotra:93b::,Rasio:93::,Peale:93::}. 
Subsequently, these studies were used to confirm the existence of the \psr 
planetary system through detecting, in the timing observations of the pulsar,
the presence of those non-keplerian variations \citep{Wolszczan:94::}

However, the application of non-keplerian dynamics goes further than
the confirmation of the discovery.  It can be used to derive some
interesting information about the system which is not otherwise
accessible.  The aim of this paper is to apply the theory of perturbed
planetary motion and derive an improved model for the timing
observations of PSR~B1257+12.  Such model, as we show on simulations,
should lead to determination of the masses of planets ${\cB}$ and
${\cC}$, as well as the inclinations of their orbits.  To this end, in
section \ref{sec:equation} we analyze the equations of motion in the
barycentric and Jacobi coordinate systems, which we use in the paper. 
In section \ref{sec:timing} we show how these two slightly different representations
are related to the commonly used timing model for pulsars with
companions. In section \ref{sec:oscul} we demonstrate how to express 
such timing formula in terms of the osculating orbital elements. 
In section \ref{sec:semi} we show how to obtain the osculating elements of ${\cB}$
and ${\cC}$.  In section \ref{sec:timfor} we present an improved timing model
describing the motion of this system and finally, in section 
\ref{sec:tests}, we perform numerical tests which show how it can be 
used in practice.

\section{Equations of Motion}

\label{sec:equation}

Let us consider a system consisting of a neutron star $P_0$ with the 
mass $m_0$, and $N$ planets $P_i$ with masses $m_i$. In an arbitrary 
inertial reference frame equations of motion of this system have the form
\begin{equation}
\label{eq:nb}
m_i \ddot{\bR}_i = -G\sum_{ {\stackrel{\scriptstyle j=0}{j\neq i}}}^N
\frac{ m_i m_j}{R_{ij}^3}\left(\bR_i-\bR_j\right),
\end{equation}
where
\[
R_{ij}=\Vert\bR_i-\bR_j\Vert, \qquad i, j = 0, \ldots, N,
\]
and $G$ is the gravitational constant. The Hamiltonian function for system 
\mref{eq:nb} has the form
\begin{equation}
\label{eq:h}
H = \frac{1}{2}\sum_{i=0}^N\frac{1}{m_i}\bP_i^2 - \sum_{0\leq i<j\leq N}
\frac{G m_im_j}{R_{ij}},
\end{equation}
and equations \mref{eq:nb} can be written as Hamilton's equations
\begin{equation}
\label{eq:he}
\dt{\bR_i} = \pder{H}{\bP_i}, \qquad \dt{\bP_i} = -\pder{H}{\bR_i}, 
\qquad i = 0, \ldots, N.
\end{equation}
Analytical perturbation theory for a planetary system is usually formulated 
in the so-called Jacobi coordinates $\br_i$, $i=0, \ldots, N$ which are 
defined in the following way
\begin{equation}
\label{eq:jac}
\br_k = \bR_k - \frac{1}{\mu_{k-1}}\sum_{i=0}^{k-1} m_i\bR_i,\quad\text{for}
\quad k=1,\ldots, N,
\end{equation}
and 
\begin{equation}
\label{eq:jac0}
\br_0 =  \frac{1}{\mu_{N}}\sum_{i=0}^{N} m_i\bR_i,\qquad \mu_k=
\sum_{i=0}^km_i.
\end{equation}
In other words, $\br_i$ is the radius vector from the
center of mass of bodies $P_0, \ldots, P_{i-1}$ to body $P_i$, and $\br_0$ 
is the center of mass of the system. The above formulae define a one-to-one 
relationship between Cartesian inertial coordinates and Jacobi coordinates. 
The inverse relationship has the form
\begin{equation}
\label{eq:jaci}
\bR_k = \br_0 + \frac{\mu_{k-1}}{\mu_k}\br_k - \sum_{i=k+1}^N\frac{m_i}{\mu_i}
\br_i, \qquad k=0,\ldots, N,
\end{equation}
where we assume $\mu_{-1}=0$.
In terms of canonical Jacobi coordinates, Hamiltonian \mref{eq:h} reads
\begin{equation}
\label{eq:hj}
\begin{array}{l}
\displaystyle
H = \frac{1}{\mu_N}\bp_0^2 + \sum_{i=1}^N\frac{\mu_i}{\mu_{i-1}m_i}\bp_i^2
  - G\sum_{i=1}^{N}\frac{m_0m_i}{r_i} \; + \\[0.5cm]
\displaystyle
\qquad + \;\tilde H_1\left(\br_1,\ldots, \br_N\right),
\end{array}
\end{equation}
where $\tilde H_1=\tilde H_1(\br_1,\ldots, \br_N)$ is its perturbative part. 
If we assume that masses of planets are small and all 
are of the same order $\epsilon$, then $\tilde H_1$ is of order 
$\epsilon^2$, i.e., $\tilde H_1 = H_1 + {\cal O}(\epsilon^3)$, where
\begin{equation}
\label{eq:h1}
H_1 = -G\!\!\!\sum_{1\leq i< j\leq N} \!\!\!m_i m_{j}\left[ \frac{1}{r_{ij}} 
- \frac{\br_i\cdot\br_j}{r_i}\right].
\end{equation}
From the form of  Hamiltonian \mref{eq:hj} it follows that $\bp_0$ is a first 
integral and that the equations for variables 
$\{\br_1,\ldots, \br_N, \bp_1, \ldots, \bp_N\}$ do not depend on its 
particular value. Thus, we can assume that $\bp_0=\bzero$ implying 
that $\br_0$ is constant and can be set to $\br_0=\bzero$. 
This is equivalent to the assumption that our inertial frame is a certain 
barycentric reference frame. The dynamics of the system is governed by 
Hamilton's equations with the Hamiltonian of the form  
\begin{equation}
\label{eq:hper}
H = H_0 + H_1 + \cdots,
\end{equation}
where 
\begin{equation}
\label{eq:h0}
H_0 = \sum_{i=1}^N\frac{\mu_i}{\mu_{i-1}m_i}\bp_i^2  - G\sum_{i=1}^{N}
\frac{m_0m_i}{r_i},
\end{equation}
is the unperturbed part of the Hamiltonian describing a system of $N$ 
independent planets. It follows from the form of Eq. \mref{eq:h0} that 
each planet moves in a Keplerian orbit in the same way as a body with the mass 
$m_i\mu_{i-1}/\mu_i$ around a fixed gravitational center $m_0 m_i$. 
Each of these Keplerian motions can be parameterized by Keplerian elements 
$\{ T_p, a,e,i,\omega,\Omega\}$---the time of pericenter, semi-major axis, 
eccentricity, inclination, argument of pericenter  and the longitude of
ascending node, respectively.

\section{Timing model and coordinate systems}

\label{sec:timing}

Timing observations of pulsars represent measurements of the times of arrival
of pulsars pulses (TOAs). An extraordinary precision of timing measurements 
allows a detection of very low-level effects in timing residuals 
\citep[see for review][]{Lyne:98::}. In the case of a binary pulsar the 
observed TOAs exhibit periodic variations resulting from the motion of the 
pulsar around the center of mass. Such variations are modeled with the 
formula
\begin{equation}
\Delta t = x\left[\left(\cos E - e\right)\sin\omega + \sqrt{1 - e^2}\sin
E\cos\omega\right],
\end{equation}
where
\[
x = a\sin i/c, \quad E - e \sin E = n\left(t- T_p\right), \quad
n = \frac{2\pi}{P}, 
\]
and $\Delta t$ is the additional delay/advance in TOAs, $E$ is eccentric
anomaly, ${a,e,\omega,\sin i,P,T_p}$ are Keplerian elements of the orbit
and $c$ is the speed of light. When a pulsar has $N$ planets the
TOA variations become
\begin{equation}
\begin{array}{l}
\label{eq:kep::}
\displaystyle
\Delta t = \sum_{j=1}^{N}x_j\Bigl[\left(\cos E_j - e_j\right)\sin\omega_j 
\; + \\[0.5cm]
\displaystyle
\hspace{2cm} + \; \left.\sqrt{1 - e_j^2}\left.\sin E_j\cos\omega_j
\right)\right].
\end{array}
\end{equation}
Note that in practice the parameters of the model which we  
determine, by means of the least-squares fit to the data, are 
${x_j,e_j,\omega_j,P_j,T_{pj}}$, i.e., the projection of  semi-major axis
of the {\it pulsar's} orbit, 
eccentricity, argument of pericenter, orbital period and time of pericenter. 
In order to precisely understand and interpret these parameters we describe 
the pulsar's motion in the barycentric reference frame with
the $z$-axis of the system directed toward the barycenter of the solar system 
and $xy$ plane of the reference frame in the plane of the sky. This way, 
we have
\begin{equation}
\label{eq:dt}
  \Delta t = -\frac{1}{c}  \bR_0\cdot \widehat\bZ ,  
  \quad \bR_0 = - \frac{1}{m_0} \sum_{i=1}^N m_i\bR_i,
\end{equation}
where $\widehat\bZ$ is the unit vector along the $z$-th axis of the pulsar
system barycentric frame and $\bR_j$ are barycentric positions of planets. 
Assuming that
\begin{equation}
\frac{m_j}{m_0}a_j\sin i_j = x_j c, 
\quad \frac{Gm_0}{(1 + m_j/m_0)^2} = n_j^2 a_j^3,
\end{equation}
with {\it planets'} orbital parameters $a_j,n_j,e_j,\omega_j,T_{pj}$, 
we can most naturally interpret the motion of the pulsar as a superposition 
of the elliptic motions of its planets around the barycenter of the system. 
However, as it was mentioned in the previous section, the analytical 
perturbation theory is usually formulated in Jacobi coordinates in which 
the TOA variations become
\begin{equation}
\Delta t = -\frac{1}{c}  \bR_0\cdot \widehat\bZ, \quad  
\bR_0 = -  \sum_{j=1}^N \kappa_j\br_j, \qquad
\kappa_j = \frac{m_j}{\mu_j}, \quad
\end{equation}
where $\br_j$ are positions of planets. Furthermore, we have the following 
relations
\begin{equation}
\kappa_ja_j\sin i_j = x_j c, \quad 
\frac{Gm_0\mu_j}{\mu_{j-1}} = \frac{Gm_0}{1 - \kappa_j}  
= n_j^2 a_j^3. 
\end{equation}
Thus from the timing formula for TOA variations of a pulsar with planets
it is possible to obtain two somewhat different descriptions of the pulsar
motion. Although, they both represent a sum of a certain number of elliptic 
motions, the interpretation of some of their parameters is slightly 
different. Throughout the rest of this paper, we will use Jacobi coordinates 
as they are more convenient in the formulation of the theory of perturbed 
motion.

\section{Osculating orbital elements}
\label{sec:oscul}
The non-keplerian motion of the \psr system can be described by 
means of the osculating ellipses (i.e. by means of ellipses which
parameters change with time). The time evolution of orbital elements
can be determined by solving the classical Lagrange's perturbation
equations \citep[see, for example][]{Brouwer:61::} with the perturbation
Hamiltonian given by equation \mref{eq:h1}. Such approach leads
to the solution in the form 
\begin{equation}
\label{eq:dx}
\cx = \cx^0 + \Delta\cx\left(t-t_0\right),
\end{equation}
where $\cx$ stands for a specific orbital element, $\cx^0$ its initial 
value and $\Delta\cx$ for its time dependent part of small magnitude 
$\Delta\cx(t-t_0)/\cx^0 \ll 1$. In the case of the \psr planetary
systems the most significant part of the perturbations comes from
 planets $\cB$ and $\cC$. Therefore, we 
can assume that the orbital elements of  planet $\cA$ are approximately 
constant while the elements of planets $\cB$ and $\cC$ change with time.
Thus using the formulae for $\Delta t$ from the previous section 
(Eq. \mref{eq:kep::}) and assuming the time evolution of orbital elements in 
the form \mref{eq:dx}, the additional TOA variations 
$\delta t_j, j = \{\cB,\cC\}$ due to the interactions between 
planets $\cB$ and $\cC$ can be expressed as follows
\begin{equation} 
\label{eq:small}
\begin{array}{l}
\displaystyle
\frac{c\delta t_j}{\kappa_j\sin i^0_j} = 
-\Delta h_j\left(\frac{3}{2}a_j^0 + \frac{1}{2}a_j^0
\cos\left(2\lambda^0_j\right)\right) \; +
\\[0.3cm] 
\displaystyle
\hspace{2cm} + \; \frac{1}{2}\Delta k_j a_j^0\sin\left(2\lambda^0_j\right) 
\; + \\[0.3cm]
\displaystyle
\hspace{2cm} + \; \Delta a_j\sin\lambda_j^0 + 
\Delta\lambda_ja_j^0\cos\lambda^0_j,
\end{array} 
\end{equation}
where
\begin{equation}
\begin{array}{l}
\displaystyle
h_j = e_j\sin\omega_j, \quad k_j = e_j\cos\omega_j, \\[0.3cm]
\displaystyle
\lambda_j = n_j(t - T_{pj}) + \omega_j,
\end{array}
\end{equation} 
and equation \mref{eq:small} is given to first order in 
$\Delta a_j, \Delta\lambda_j, \Delta h_j, \Delta k_j$ and
the lowest order in $e_j$ (or $h_j$ and $k_j$, since
the eccentricities of planets $\cB$ and $\cC$ are very small).
The above equations can be obtained from the following 
well-known expansions
\begin{equation}
\begin{array}{c}
\displaystyle
\cos E = -\frac{e}{2} + 2\sum_{k \in {\cal Z}_0}\frac{1}{k}J_{k-1}(ke)\cos(k M),
\\[0.3cm]
\displaystyle
\sin E = \frac{2}{e}\sum_{k \in {\cal Z}_0}\frac{1}{k}J_{k-1}(ke)\sin(k M),
\quad M = \lambda - \omega
\end{array}
\end{equation}
where $J_k(x)$ is a Bessel function of the first kind of order $n$ and
argument $x$, ${\cal Z}_0$ denotes the set of all positive and negative integers  
excluding zero; and an approximate relation for Bessel functions of 
small arguments
\begin{equation}
J_{k}(x) \approx \frac{x^k}{2^k k!}, \quad x << 1
\end{equation}

Now, our next step is to find the explicit form of the functions
\begin{equation}
\begin{array}{l}
\displaystyle
a_j(t) = a_j^0 + \Delta a_j(t), \quad \lambda_j(t) = \lambda_j^0 + 
\Delta \lambda_j(t), \\[0.3cm]
\displaystyle
h_j(t) = h_j^0 + \Delta h_j(t), \quad k_j(t) = k_j^0 + \Delta k_j(t).
\end{array}
\end{equation}

\section{Semi-analytical perturbation theory}

\label{sec:semi}
Mutual gravitational interactions between planets cause periodic and 
secular changes of their orbital elements. In the case of  \psr
 periodic variations can be related to conjunctions (`close encounters') 
of  planets $\cB$ and $\cC$ with the frequency $n_{ce} = n_\cB - n_\cC$
and to the 3:2 near-commensurability with the frequency 
$n_{r} = 3n_{\cC}-2n_{\cB}$. The dynamics of this system was
studied by \citet{Rasio:92::} and \citet{Malhotra:93a::} \citep[see 
also][]{Malhotra:93b::,Rasio:93::,Peale:93::}. \citet{Malhotra:93a::} 
solved the Lagrange's perturbation equations
for the orbital elements assuming non-resonant system with coplanar orbits. 
This first order perturbation theory is valid for $(\sin i)^{-1} \gtrsim 10$ 
(orbital inclinations larger than about 6 degrees) which corresponds to mass 
ratios $m_j/m_0 \gtrsim 6\times10^{-5}$ (and non-resonant case). When  
planets $\cB$ and $\cC$ are locked in the exact resonance it is necessary 
to develop a different theory of motion \citep{Malhotra:92::,Rasio:93::}.

It turns out that the first-order perturbation theory for this system
developed by \citet{Malhotra:93a::} is not accurate enough for 
the purpose
of determination of the planets' masses. Therefore, in this paper we present
a new approach to this problem which precisely addresses the issue.
Namely, first we assume that the orbits of ${\cB}$ and ${\cC}$ are not
coplanar however their relative inclination $I$ is small. The geometry
of the system can be represented as in Fig. 1 and the perturbative part of
Hamiltonian expanded to the first order in $I$ and the product of the
mass of planets ${\cB}$ and ${\cC}$, $m_1m_2$, has the following form
\begin{equation}   
H_1 = -\frac{Gm_1m_2}{r_2}\left[\left(1 - 2\frac{r_1}{r_2}
\cos\psi 
+ \left(\frac{r_1}{r_2}\right)^2\right)^{-1/2} -\frac{r_1}{r_2}\cos\psi\right],
\end{equation}
where
\[
\begin{array}{l}
\displaystyle
\psi = f_1 + \omega_1 - f_2 - \omega_2 - \tau, \quad 
\tau = \tau_1 - \tau_2 ,\\[0.5cm]
\displaystyle
r_j = \frac{a_j\left(1 - e_j^2\right)}
{1 + e_j\cos f_j}, \quad j={1,2},
\end{array}
\]
and $f_j$ is the true anomaly. Note that in the above expansion,
$H_1$ does not depend on $I$. The first term in which $I$ appears is 
of the second order; precisely it is $\sin^2(I/2)$. Thus as long as
$\sin^2(I/2)$ is negligible, the above expansion is valid.

Next, from the classical form of Lagrange's perturbation equations 
\citep[see, for example][]{Brouwer:61::} we can obtain the following set 
of the first order differential equations for the elements $a_j, e_j, 
\omega_j, \lambda_j$
\begin{equation}
\label{osc}
\begin{array}{c}
\displaystyle
\dot a_j = \frac{-2}{\mu_jn_ja_j}
\frac{\partial H_1}{\partial\sigma_j}, \\[0.7cm]
\end{array}
\end{equation}
\[
\begin{array}{c}
\displaystyle
\dot e_j = \frac{1}{\mu_jn_ja_j^2e_j}\left[
-\left(1 - e_j^2\right)\frac{\partial H_1}{\partial\sigma_j}
+ \sqrt{1 - e_j^2}\frac{\partial H_1}{\partial\omega_j}\right], \\[0.7cm]
\displaystyle
\dot\omega_j = 
 \frac{-\sqrt{1 - e_j^2}}{\mu_jn_ja_j^2e_j}
\frac{\partial H_1}{\partial e_j}, \\[0.7cm]
\displaystyle
\dot\lambda_j = n_j + \frac{2}{\mu_jn_ja_j}\frac{\partial H_1}{\partial a_j}
- \frac{e_j\sqrt{1 - e_j^2}}{\mu_jn_ja_j^2\left(1 + \sqrt{1 - e_j^2}\right)}
\frac{\partial H_1}{\partial e_j},
\end{array}
\]
where $n_j$ is the mean motion and $\sigma_j$ is related to time of
pericenter through $\sigma_j = -n_j T_{pj}$ so as the Kepler equation
reads 
\begin{equation}
E_j - e_j\sin E_j = n_jt + \sigma_j = \lambda_j - \omega_j,
\end{equation}
and the true anomaly $f_j$ is related to the eccentric anomaly $E_j$
through 
\begin{equation}
\tan\frac{f_j}{2} = \sqrt{\frac{1 + e_j}{1 - e_j}}\tan\frac{E_j}{2}.
\end{equation}
And it should be noted that the elements $\Omega_j, i_j$ remain approximately
constant in the case of small relative inclinations. 

In principle, such set of equations could be solved for $a_j, e_j, \omega_j,
\lambda_j$. In practice however, it is extremely complicated. On the other 
hand, in fact we do not need analytical formulae such as those presented in 
\citet{Malhotra:93a::}. Thus the most suitable approach is to solve 
this problem numerically. In order to do so we just need the explicit form
of the right-hand-side functions of  equations \mref{osc} which can be
easily obtained using the perturbing Hamiltonian $H_1$. Subsequently, 
the equations can be solved numerically for $a_j, e_j, \omega_j, \lambda_j$. 
As we show in section 7, this approach gives very accurate results, much 
more accurate than any reasonable analytical treatment. From the form of 
equations \mref{osc} it follows that the change of  orbital parameters 
$\Delta\cx$ (strictly speaking here $\Delta\cx$ stands for 
$\Delta\lambda,\Delta h,\Delta k$ and $\Delta a/a_0$) is proportional to 
$m_1/m_0$ and $m_2/m_0$. Precisely, corrections $\Delta\cx_1$ are proportional
to $m_2/m_0$ and $\Delta\cx_2$ are proportional to $m_1/m_0$. Let us finally 
note that in this model there is one additional parameter $\tau$ which in 
general is not known a priori. However, it can be determined through a 
least-squares analysis of the data. 

\section{Timing formula}
\label{sec:timfor}

We have now all the components necessary to derive a useful timing formula
which will describe the motion of the \psr system. First of all, let us note
that due to the small mass of planet ${\cal A}$ we will have:
\begin{equation}
\begin{array}{c}
\displaystyle
\kappa_{\cA} = \kappa_1 = \frac{m_{\cA}}{m_0}, \\[0.4cm]
\displaystyle
\kappa_{\cB} = \kappa_2 = \frac{m_{\cB}}{m_0 + m_{\cA} + m_{\cB}}
\approx \frac{m_{\cB}}{m_0 + m_{\cB}}, \\[0.4cm] 
\end{array}
\end{equation}
\[
\kappa_{\cC} = \kappa_3 = \frac{m_{\cC}}{m_0 + m_{\cA} + m_{\cB} + m_{\cC}}
\approx \frac{m_{\cC}}{m_0 + m_{\cB} + m_{\cC}}. 
\]

The problem of finding masses and inclinations of planets ${\cB}$ and
${\cC}$ can be now formulated as a least-squares problem in which we fit
to the data the following function
\begin{equation}
\Delta t = \Delta t_{K}(\Psi) + 
\delta t_{\cB}(\Psi,\gamma_{\cB},\gamma_{\cC}) + 
\delta t_{\cC}(\Psi,\gamma_{\cB},\gamma_{\cC}),
\end{equation}
where 
\begin{equation}
\begin{array}{l}
\displaystyle
\Psi = \left\{x^0_{\cB},n^0_{\cB},e^0_{\cB},\omega^0_{\cB},
T^0_{p\cB},x^0_{\cC},n^0_{\cC},e^0_{\cC},\omega^0_{\cC},
T^0_{p\cC}\right\}, \\[0.4cm]
\displaystyle
\gamma_{\cB} = \frac{m_{\cB}}{m_0}, \quad \gamma_{\cC} =
\frac{m_{\cC}}{m_0},
\end{array}
\end{equation}
$\Delta t_{K}(\Psi)$ describes the Keplerian part of the motion given
by  equation \mref{eq:kep::} and
\begin{equation}                                      
\begin{array}{l}
\displaystyle
\delta t_{\cB} = x_{\cB}^0\Bigg[-\Delta h_{\cB}\left(\frac{3}{2} 
+ \frac{1}{2}\cos\left(2\lambda^0_{\cB}\right)\right) \; +
\\[0.4cm] 
\displaystyle
\hspace{2cm} + \; \frac{1}{2}\Delta k_{\cB}\sin
\left(2\lambda^0_{\cB}\right) 
\; + \\[0.4cm]
\displaystyle
\hspace{2cm} + \; \frac{\Delta a_{\cB}}{a_{\cB}^0}\sin\lambda_{\cB}^0 + 
\Delta\lambda_{\cB}\cos\lambda^0_{\cB}\Bigg],\\[0.5cm]
\displaystyle
\delta t_{\cC} = x_{\cC}^0\Bigg[-\Delta h_{\cC}\left(\frac{3}{2} 
+ \frac{1}{2}\cos\left(2\lambda^0_{\cC}\right)\right) \; +
\\[0.4cm] 
\displaystyle
\hspace{2cm} + \; \frac{1}{2}\Delta k_{\cC}\sin
\left(2\lambda^0_{\cC}\right) 
\; + \\[0.4cm]
\displaystyle
\hspace{2cm} + \; \frac{\Delta a_{\cC}}{a^0_{\cC}}
\sin\lambda_{\cC}^0 + 
\Delta\lambda_{\cC}\cos\lambda^0_{\cC}\Bigg],
\end{array}
\end{equation}
describe its non-keplerian part. In this model the initial values of 
osculating orbital
elements replace the Keplerian elements as parameters of the model
and we additionally have the parameters, $\gamma_{\cB}$ and 
$\gamma_{\cC}$, related to the masses of ${\cB}$ and ${\cC}$. We also
need the derivatives of $\Delta t$ with respect to the parameters 
$\{\Psi,\gamma_{\cB},\gamma_{\cC}\}$. With sufficient accuracy the
derivatives with respect to $\Psi$ can be computed from the Keplerian part
$\Delta t_K$ of the model only and the derivatives with respect to $\gamma_{\cB}$ and 
$\gamma_{\cC}$ can be easily obtained as $\delta t_{\cB}$ and
$\delta t_{\cC}$ are proportional to them
\begin{equation}
\frac{\partial\Delta t}{\partial{\gamma_{\cB}}} = 
\frac{\delta t_{\cC}}{\gamma_{\cB}}, \quad
\frac{\partial\Delta t}{\partial{\gamma_{\cC}}} = 
\frac{\delta t_{\cB}}{\gamma_{\cC}}.
\end{equation}

Now, determination of masses of ${\cB}$ and ${\cC}$ and inclinations 
$i_{\cB}$ and $i_{\cC}$ can be carried out in the following way. First, we 
assume that the mass of the pulsar $m_0$ is canonical (i.e. $1.4$M$_{\odot}$) 
and derive $m_{\cB}$, $m_{\cC}$ directly from $\gamma_{\cB}$ and $\gamma_{\cC}$.
Subsequently inclinations can be computed from the following equations
\begin{equation}
c x^0_j = \kappa_j a^0_j\sin i^0_j,\quad
\frac{Gm_0}{1 - \kappa_j} = {n^0_j}^2{a^0_j}^3,
\end{equation}
where $j = \{\cB,\cC\}$.

\section{Numerical tests and conclusions}

\label{sec:tests}

We start the tests with the comparison of our approach for computing
osculating elements with that of \citet{Malhotra:93a::}. It turns
out that the most significant component of the non-keplerian motion comes
from the changes of mean longitude $\lambda_j$. Therefore in Fig. 2
we present $\Delta\lambda_j$ for the case of coplanar, edge-on orbits.
As we can see, the approach developed in this paper is in fact as good
as the integration of full equations of motion. For other elements
we obtain very similar results. It should be mentioned that because 
the model of \citet{Malhotra:93a::} is not accurate  enough in 
predicting the secular change of $\Delta\lambda_j$ and because
$\Delta\lambda_j$ are proportional to $\gamma_{\cB}$ and $\gamma_{\cC}$, using 
such model would result in a significant error in determination of the 
planets' masses. Next, we compared the TOA residuals due to the non-keplerian
part of the motion calculated from our model with those obtained by means
of  numerical integration of full equations of motion. In Fig. 3,
as one can see, for small relative inclinations (Fig. 3 (a) and (b)) our 
model is very accurate. For larger $I$ (Fig. 3 (c) and (d)) one can 
see small differences but this is consistent with the assumptions made 
to obtain the model.
  
Finally, we performed the tests which show how the derived timing model
can be used to obtain the masses of planets. To this end we used the 
TEMPO software package (see the Internet location 
{\tt http://pulsar.princeton.edu/tempo}) modified to include our model
of the non-keplerian motion of  \psr. We simulated two different
sets of artificial TOAs assuming the parameters of  \psr
\citep{Wolszczan:94::}. The first set of TOAs sampled every day 
covered a period of 10 years. We also added Gaussian noise with
$\sigma = 0.1 \mu s$. The second set was prepared to resemble the real
timing observations of PSR B1257+12. Under such conditions we simulated
TOAs for the cases described in Fig. 3. Subsequently, we applied
the modified TEMPO to obtain the masses of planets $\cB$ and $\cC$. The
results are presented in Table 1.

The tests performed on  Set 1 were used to establish a limit on 
applicability of our model. As one can see, for the relative inclinations
$I$ of about 10 degrees the model becomes inaccurate and the relative error
of determination of masses is at the level of 20$\%$. Because the relative
inclination weakens the interactions of planets, using the model which
assumes a small $I$, results, in such situations, in the determined masses 
that are smaller than the real ones. From the tests performed on  Set 2 
we also learn that this error is bigger than uncertainty originating in
TOA measurement errors. On the other hand, when $I$ is relatively small 
we obtain a very accurate determination of  masses thus proving that
our model can be successfully applied as long as the assumptions made
to derive it are satisfied.

To sum up, our model can be applied in all
the cases when the `observational' uncertainties of the planets' masses
are bigger than the uncertainties resulting from the assumptions made to
derive the timing model. This, in general, depends on  masses of planets
and the relative inclination of the orbits $I$ as well as the
characteristic of the observations  but one can estimate that
in the case of the \psr data $I$ should be smaller than 10 degrees.
We should also mention that for non-coplanar orbits the angle
$\tau$ will be in general different from zero therefore we have to find
it in order to get the proper values of the planets' masses. It can 
be done by computing the least-squares value of $\chi^2$ for a range of 
$\tau$ and then choosing the $\tau$ that corresponds to the minimum
of $\chi^2$. And eventually, one has to remember that the \psr planetary
system could be in such configuration that a more significant alteration
must be done to our model. First of all, the relative inclination
of the orbits could be large. In such  case one would have to use
the full form of the perturbative Hamiltonian $H_1$ which would lead
to a much more complicated model with four additional parameters $\Omega_{\cB},
\Omega_{\cC},i_{\cB},i_{\cC}$ (instead of one $\tau$). Secondly, the 
presence of a massive distant planet, if significant for the theory of 
motion, would have to be taken into account.

\clearpage

%
%

\figcaption[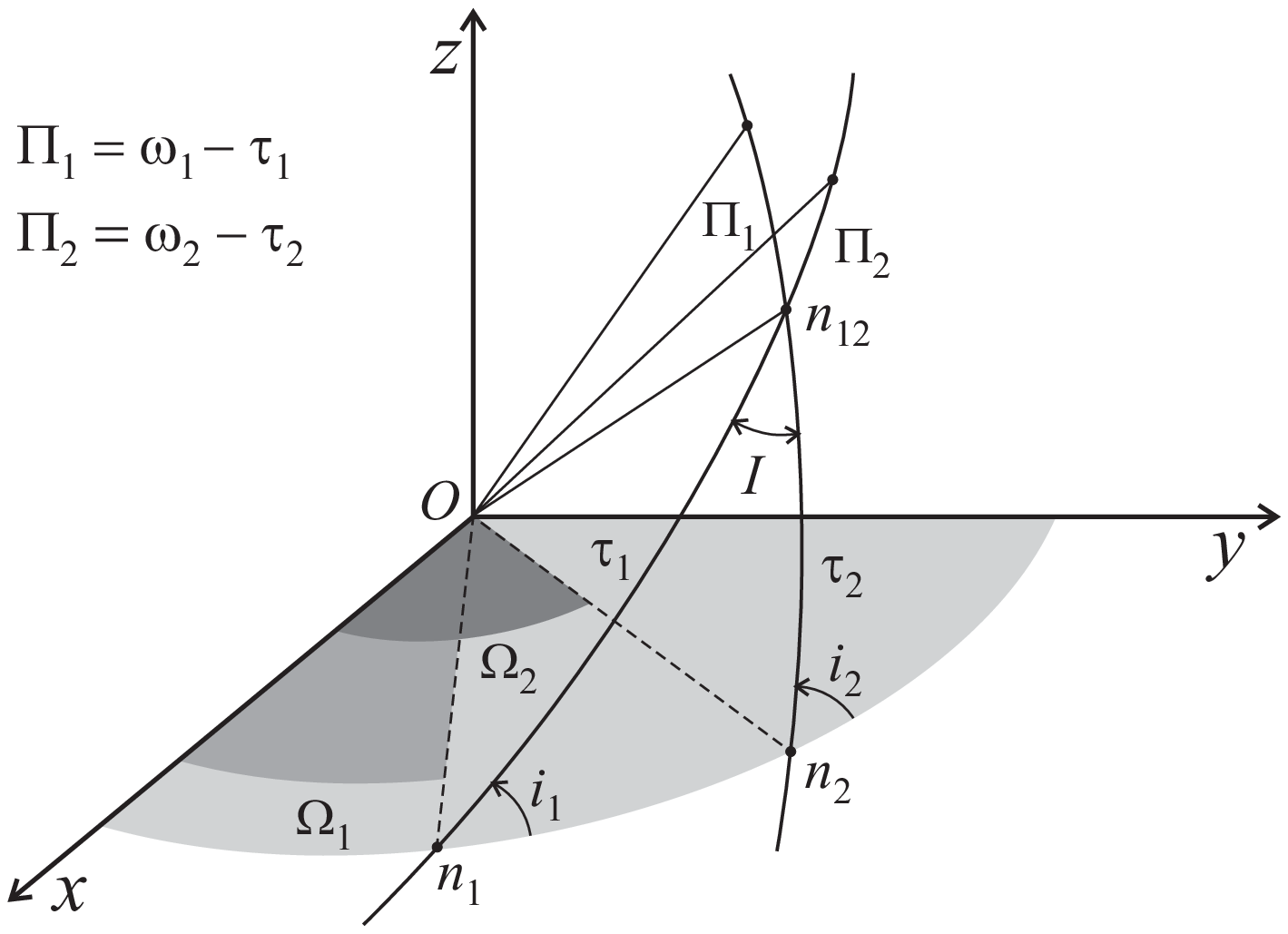]
{Geometry of the system. The angles $\omega_1$, $\omega_2$ 
are the arguments of pericenter, $\Omega_1$, $\Omega_2$ longitudes 
of ascending node and $i_1$, $i_2$ inclinations of the orbit of the 
planets ${\cB}$ and ${\cC}$; $\tau_1$ and $\tau_2$ are the angles
$n_1On_{12}$ and $n_2On_{12}$ respectively. The angles $I,\tau_1,\tau_2$
can be found by solving the spherical triangle $n_1n_{12}n_2$.}

\figcaption[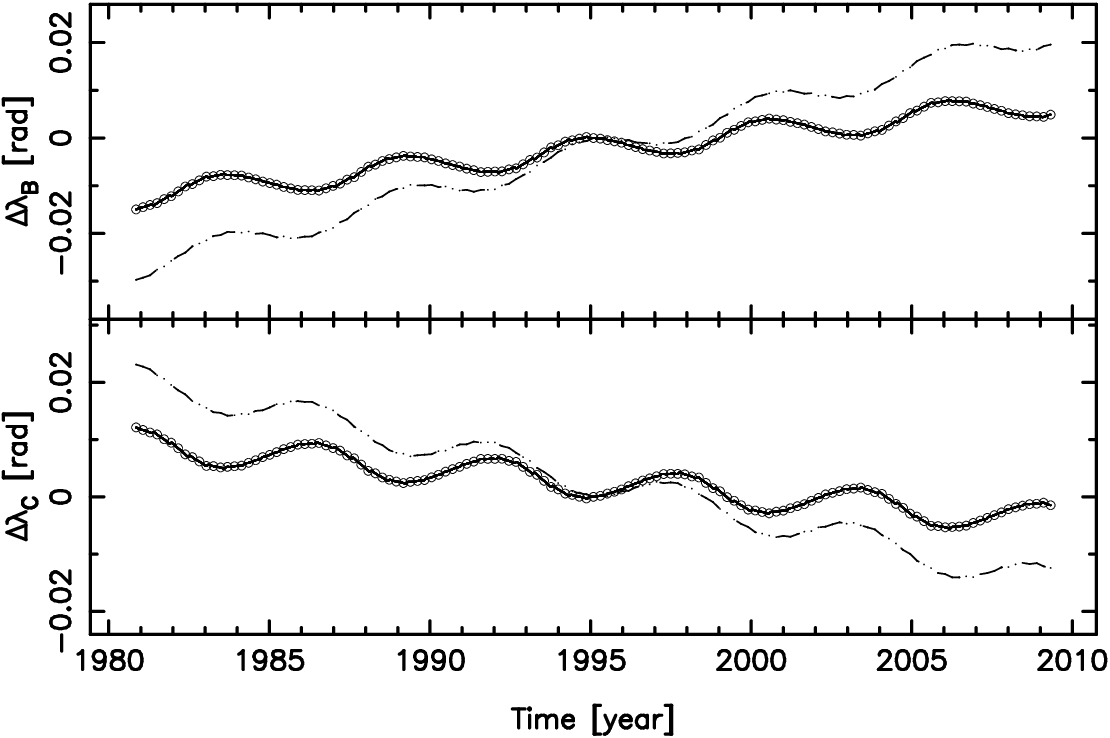]
{Time evolution of the mean longitude 
$\Delta\lambda(t) = \lambda(t) - \lambda^0(t)$ for 
the planets $\cB$ and $\cC$ in the case of coplanar, edge-on orbits. 
The solid line indicates the solution
by means of the numerical integration of full equations of motion,
the open circles indicate the solution obtained with the approach
presented in this paper and the dash-dotted line with the model by
\cite{Malhotra:93a::}.}

\figcaption[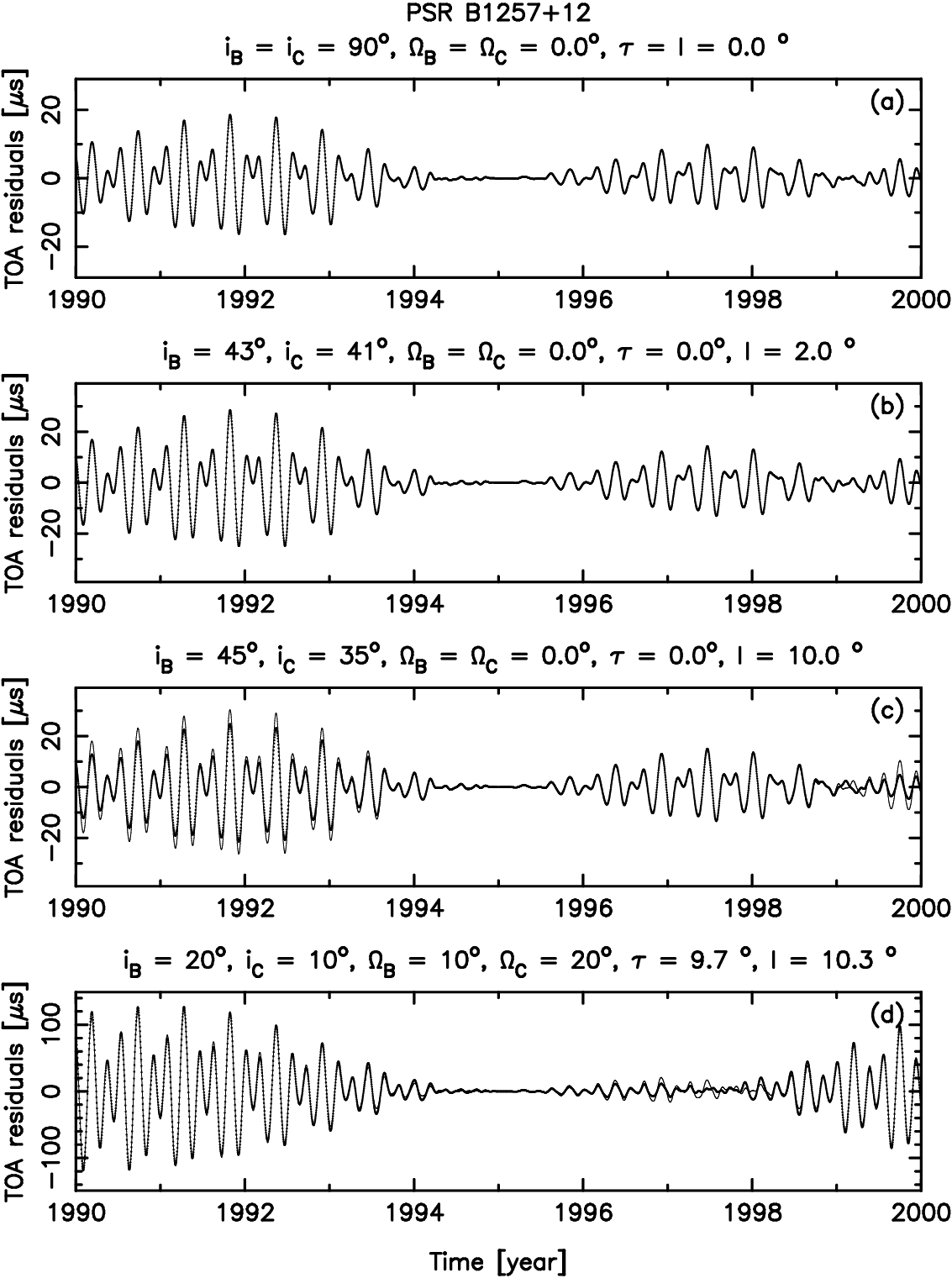]
{TOA residuals due to the non-keplerian part of motion of the 
PSR B1257+12 in four different configurations. The solution
obtained by means of the numerical integration of full equations
of motion is indicated with the dots and the one obtained using
our model with the solid line.}

\clearpage

%
%

%
%

\begin{deluxetable}{lrrrrrrrrrr}
\tablecolumns{11}
\tablewidth{0pc}
%
\tablecaption{ Assumed and derived parameters of simulations
\tablenotemark{a}}
\tablehead{
\colhead{} &
\multicolumn{4}{c}{Assumed} &
\colhead{} &
\multicolumn{2}{c}{Set 1} &
\colhead{} &
\multicolumn{2}{c}{Set 2} \\
\cline{2-5} \cline{7-8} \cline{10-11}\\
\colhead{} &
\colhead{$\tau$} &
\colhead{$I$} &
\colhead{$m_{\cB}$} &
\colhead{$m_{\cC}$} &
\colhead{} &
\colhead{$m_{\cB}$} &
\colhead{$m_{\cC}$} &
\colhead{} &
\colhead{$m_{\cB}$} &
\colhead{$m_{\cC}$} \\
\colhead{} & \colhead{} & \colhead{} & 
\colhead{$[M_{\earth}]$} & \colhead{$[M_{\earth}]$} &
\colhead{} &
\colhead{$[M_{\earth}]$} & \colhead{$[M_{\earth}]$} &
\colhead{} &
\colhead{$[M_{\earth}]$} & \colhead{$[M_{\earth}]$} }
\startdata
(a)\dotfill& $0.\!\!^\circ0$ & $0.\!\!^\circ0$ & 3.41 & 2.83 & 
& 3.42(1) & 2.82(1) & &
3.53(51) & 2.63(51) \\
(b)\dotfill& $0.\!\!^\circ0$ & $2.\!\!^\circ0$ &4.99 & 4.32 &   
& 4.98(1) & 4.26(1) & & 
5.07(51)& 4.06(48) \\
(c)\dotfill& $0.\!\!^\circ0$ & $10.\!\!^\circ0$& 4.82 & 4.94 &  
& 4.06(3) & 4.11(3) &  &
4.06(51)&3.79(48) \\
(d)\dotfill& $9.\!\!^\circ7$ & $10.\!\!^\circ3$ &9.96 & 16.33 &  
& 8.42(12) & 13.39(12)& &
8.25(45)&13.16(42) \\
\enddata
\tablenotetext{a}{Figures in parentheses are $3\sigma$
uncertainties in the last digits quoted.}
\end{deluxetable}

\clearpage

%
%

%
%

\begin{figure}
\figurenum{1}
\epsscale{0.75}
\plotone{fig1.ps}
\caption{}
\end{figure}
%

%
%

\begin{figure}
\figurenum{2}
\epsscale{0.9}
\plotone{fig2.ps}
\caption{}
\end{figure}
%

%
%

\begin{figure}
\figurenum{3}
\epsscale{0.9}
\plotone{fig3.ps}
\caption{}
\end{figure}

\end{document}